\documentstyle[12pt,aaspp4]{article}

\newcommand{\co}{{\rm $^{12}$CO}}

\newcommand{\cn}{{\rm C$^0$}}
\newcommand{\cp}{{\rm C$^+$}}
\newcommand{\ci}{\mbox{\rm [\ion{C}{1}]}}
\newcommand{\oi}{\mbox{\rm [\ion{O}{1}]}}
\newcommand{\cii}{\mbox{\rm [\ion{C}{2}]}}

\newcommand{\hii}{\mbox{\rm \ion{H}{2}}}
\newcommand{\jone}{($J=1\rightarrow0$)}
\newcommand{\jtwo}{($J=2\rightarrow1$)}

\newcommand{\fsone}{($^3$P$_1\rightarrow^3$P$_0$)}
\newcommand{\fstwo}{($^3$P$_2\rightarrow^3$P$_1$)}
\newcommand{\fseno}{($^3$P$_0\rightarrow^3$P$_1$)}
\newcommand{\fsowt}{($^3$P$_1\rightarrow^3$P$_2$)}
\newcommand{\fscii}{($^2$P$_{3/2}\rightarrow^2$P$_{1/2}$)}
\newcommand{\percmcu}{cm$^{-3}$}

\newcommand{\kmpers}{km~s$^{-1}$}
\newcommand{\Kkmpers}{K~km~s$^{-1}$}
\newcommand{\intunits}{erg~s$^{-1}$~cm$^{-2}$~sr$^{-1}$}

\newcommand{\Ico}{\mbox{\rm I$_{\rm CO}$}}
\newcommand{\Ici}{\mbox{\rm I$_{\rm [CI]}$}}
\newcommand{\Icii}{\mbox{\rm I$_{\rm [CII]}$}}
\newcommand{\Ifir}{\mbox{\rm I$_{\rm FIR}$}}
\newcommand{\citoco}{\Ici/\Ico}
\newcommand{\ciitoco}{\Icii/\Ico}

\newcommand{\av}{\mbox{\rm A$_{\rm V}$}}
\newcommand{\xuv}{\mbox{$\chi_{\rm uv}$}}

\begin{document}

\title{First Detection of Submillimeter \ci\ Emission in the Small Magellanic Cloud}

\author{Alberto D. Bolatto\altaffilmark{1}, James M. Jackson\altaffilmark{1},
Kathleen E. Kraemer\altaffilmark{1},}
\affil{Institute for Astrophysical Research, Department of Astronomy,} 
\affil{Boston University, 725 Commonwealth Ave., Boston MA 02215} 
\and
\author{Xiaolei Zhang\altaffilmark{2}} 
\affil{Smithsonian Astrophysical Observatory, 60 Garden St. MS 78, 
Cambridge, MA 02138} 

\altaffiltext{1}{bolatto, jackson, kraemer@bu-ast.bu.edu}
\altaffiltext{2}{Currently at Raytheon ITSS and NASA GSFC, code 685, 
Greenbelt, MD 20771 (xzhang@specs1.gsfc.nasa.gov)}

\begin{abstract}
We report the first detection of \ci\ \fsone\ emission at 609 $\mu$m
in a region of the Small Magellanic Cloud (N27). Environments poor in
heavy elements and dust such as the SMC are thought to be dominated by
photodissociation regions. This is the lowest metallicity source where
submillimeter neutral carbon emission has been detected. Studying the
\citoco\ ratio in several sources spanning more than an order of
magnitude in metallicity, $Z$, we find that the \citoco\
ratio increases for decreasing $Z$. The existence of such a trend points to a
photodissociation origin for most of neutral carbon in molecular clouds,
in agreement with standard PDR models. We also report ISO FIR
spectroscopic observations of N27, and use them to derive its physical
properties. Comparison between the density and radiation field
revealed by FIR diagnostics ($n\sim300$---1000~\percmcu,
$\xuv\sim30$---$100\chi_0$) and those derived from millimeter and
submillimeter data ($n\sim10^5$~\percmcu, $\xuv\lesssim30\chi_0$)
suggests that the FIR lines originate in more diffuse gas,
and are perhaps dominated by the interclump medium. Regardless of the
cause, analysis of the FIR and mm-submm data produces a discrepancy of 
two orders of magnitude for the density of this source.
\end{abstract}


\section{Introduction}
\label{intron27}

Studies of the interstellar medium (ISM) in the Small Magellanic Cloud
(SMC) allow us to investigate star forming clouds that are poor in
heavy elements. The SMC is indeed an excellent choice to study this
phenomenon, because it is nearby ($D\sim61$~kpc; Laney \& Stobie
1994), has an unobscured line of sight and very small internal
obscuration, and possesses the lowest metallicity among the nearby
members of the Local Group ($Z\sim Z_\odot/10$; Dufour 1984).  Because
of their low metallicity, morphological irregularity, and active star
formation, the Magellanic Clouds resemble primeval galaxies.
Therefore, studies such as this will eventually allow us to draw
conclusions about those much more elusive targets: the high-z
primitive galaxies.

The depletion of heavy elements, and the consequent low dust-to-gas
ratio ($D/G_{\rm SMC}\sim1/8\,D/G_{\rm MW}$; Bouchet et al. 1985),
affects the ISM by enhancing photodissociation.  Because of its high
abundance (i.e., strong self-shielding) and cross-shielding with
atomic hydrogen lines, molecular hydrogen is mostly unaffected by $Z$.
The usual tracer molecules such as CO, however, are efficiently
destroyed in dust-poor systems. A much larger fraction of the
ISM in these systems is thus expected to be occupied by
photodissociation regions (PDRs; e.g., Tielens \& Hollenbach 1985;
Sternberg \& Dalgarno 1995). Consequently, we expect the intensities
of the spectral lines associated with the PDR (e.g., \cii, \ci) to be
enhanced when compared to molecular lines (e.g., \co\ \jone),
resulting in increasing \ciitoco\ and \citoco\ ratios (Bolatto,
Jackson, \& Ingalls 1999).

The immediate purpose of this work is twofold: 1) to test for the
existence of an increasing \citoco\ ratio with decreasing metallicity
$Z$ (\S\ref{disccin27}), and 2) to investigate the power of
far-infrared (FIR) spectroscopic diagnostics and compare their results
to those obtained using millimeter and submillimeter lines
(\S\ref{discpcn27}). An \citoco\ dependence on $Z$ could be useful for
determining metallicities in high-z sources using only radio
observations. The trend in \citoco\ with $Z$ is also a test for the
origin of neutral carbon inside molecular clouds. An increasing
\citoco\ ratio for decreasing $Z$ points to a PDR origin, while a flat
$\citoco$ ratio suggests that \cn\ is produced inside the CO cores by
means other than UV photons.

Our SMC source, the N27 nebula, was first
cataloged by Henize (1956). A later survey by Davies, Elliott and
Meaburn (1976) calls it DEM40. Its associated IR peak
is also known as LIRS49 (Schwering \& Israel 1989). This region was
mapped by the Swedish-ESO Submillimeter Telescope (SEST) in the \jone\
and \jtwo\ transitions of CO (Rubio et al. 1993), and studied in
detail in several molecular transitions by Heikkil\"a et
al. (1999). Albeit a weak emitter when compared to Galactic sources,
the molecular cloud associated with N27 features some of the strongest
CO \jone\ emission in the SMC (Israel et al. 1993).

\section{Observations}
\subsection{Far Infrared Data}
\label{obsfirn27}
We observed N27 using the Long Wavelength Spectrometer (LWS) onboard
the Infrared Space Observatory satellite (ISO), at
$R=\lambda/\Delta\lambda\sim200$. These LWS data are a 4160 second-long 
full-grating scan obtained on 20 December 1997 as part of the
LOMETPDR project. The instrumental characteristics are described by
Clegg et al. (1996), and the calibration of the ISO LWS instrument is
described by Swinyard et al. (1996). The data pipeline was version 8.7
and the observations were processed using ISAP version 2.0.  We
averaged the spectra across scans and scan direction, but not across
detectors, with the default wavelength spacing of 0.135 $\mu$m.  The
averaging was done after 2.5$\sigma$ clipping of outlying points. 
For the purposes of obtaining an \Ifir\ estimate only, we joined the
detectors together by offsetting them with the dark correction method in
ISAP, using detector LW1 as the reference.  Table \ref{tabiso}
summarizes the spectroscopic results.

We can assess the calibration of the LWS data by comparing it with
previous \cii\ measurements of N27 featuring
a similar beam size (Israel \& Maloney 1993). The
intensity measured by FIFI at the LWS pointing is
\Icii$\approx9\times10^{-4}$ \intunits. The LWS measured intensity
after applying the recommended correction factors for beam size and
extended source is \Icii$\approx9.5\times10^{-4}$ \intunits, in
remarkable agreement with the FIFI observations. The measurements
listed in Table \ref{tabiso} include both corrections.

\subsection{Submillimeter Data}
\label{obssubmmn27}
\begin{figure}[p]
\plotone{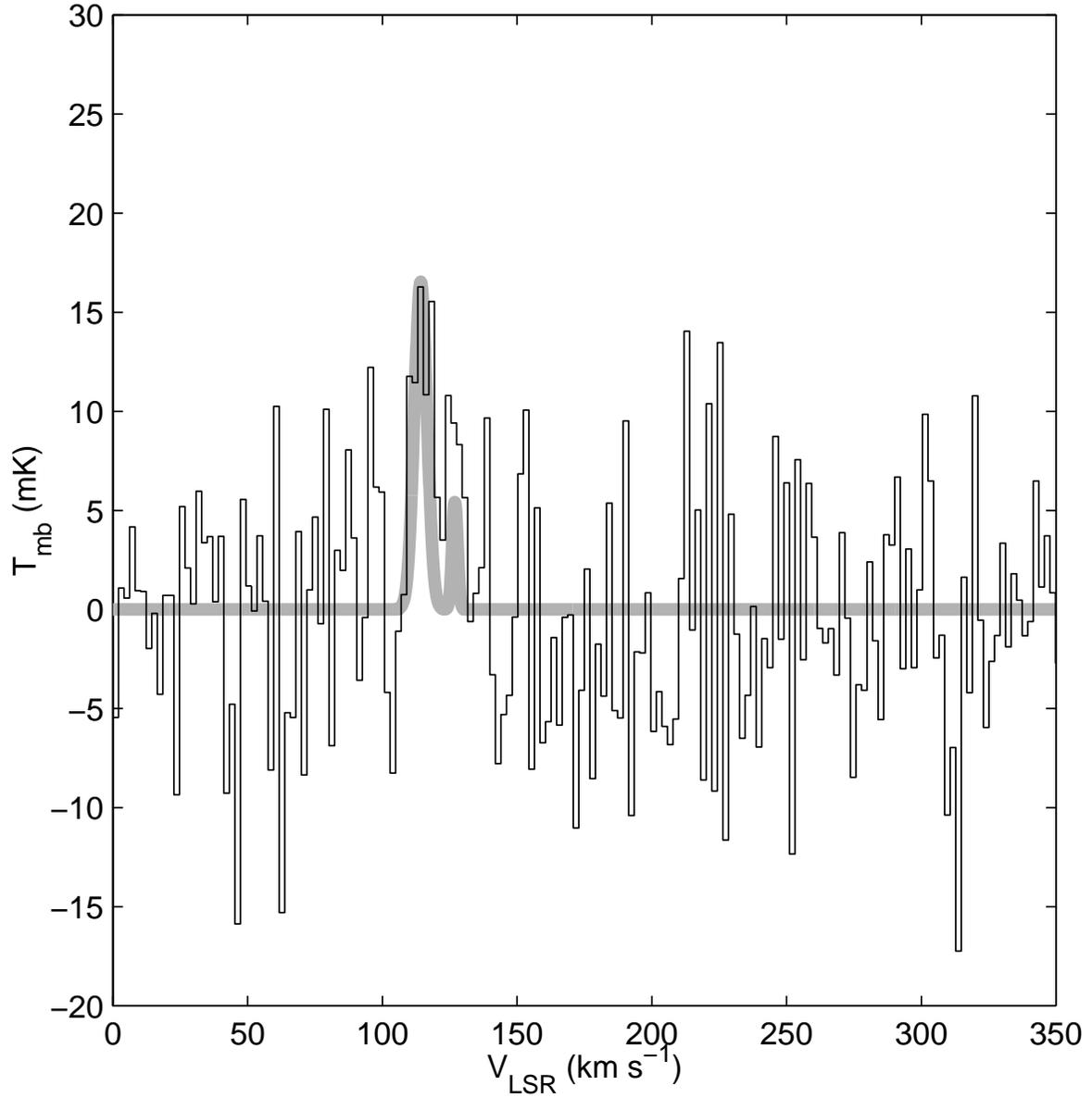} 
\figcaption[fig1.eps]{Averaged \ci\ spectrum of N27 obtained by AST/RO
(Hanning-smoothed to $\Delta v\sim2.1$ \kmpers). Superimposed are 
the rescaled Gaussian \co\ components measured by Israel et al. (1993).
There is a hint of the 127 \kmpers\ component in the \ci.\label{cispectrum}}
\end{figure}

We observed the \ci\ \fsone\ transition of neutral atomic carbon at
$\nu\simeq492.1607$ GHz (609 $\mu$m) using AST/RO, the Antarctic
Submillimeter Telescope and Remote Observatory located at the
Amundsen-Scott South Pole base (Stark et al. 1997).  The observations
were obtained on 3 to 7 July of 1998, using the AST/RO SIS
quasi-optical receiver, with a system temperature
$T_{sys}\sim1800$---2400~K. The backend was the 2048 channel low
resolution AOS.  The spectra were observed in position switching mode,
chopping 15\arcmin\ in Azimuth (same as R.A. at the
pole). At 492 GHz the telescope beam had
HPBW$\sim3.8\arcmin\pm0.3$. The forward efficiency determined from
skydips was 70\%, and it is assumed to be identical to $\eta_{mb}$.
The pointing was better than $30\arcsec$.  A $6\arcmin\times6\arcmin$
region centered on this source was mapped on a 30\arcsec\ grid,
considerably oversampling the beam.  The data were calibrated using
the standard procedure for AST/RO, which includes sky, hot and cold load
measurements every 30 minutes, and processed using the COMB
package. The total accumulated integration time, after rejecting bad
spectra, was 63 hours. The spectrum obtained after averaging all the
observations together is shown in Fig. \ref{cispectrum}.  A gaussian
fit to the data gives $v_{cen}=115.2\pm1.4$~\kmpers, $\Delta
v=9.3\pm3.3$~\kmpers, and $T_{max}=14\pm4$~mK (main beam
brightness). The location of the \ci\ line agrees very well with the
strongest CO component, for which $v_{cen}=114.5$ \kmpers\ (Rubio et
al. 1993).

The relatively weak \ci\ emission from N27, combined with its small
size and consequent beam dilution, produced a low signal-to-noise
map. Thus the analysis in \S\ref{disccin27} will be focused on the
averaged spectrum for the region.

\section{Results and Discussion}

\subsection{Metallicity and the \citoco\ Ratio}
\label{disccin27}

The prime motivation for this study is to determine the \citoco\
intensity ratio in the SMC, an environment that because of its low
metallicity and low dust-to-gas ratio should be dominated by PDRs
(cf., \S\ref{intron27}). To do this we compare our \ci\ data with the
\co\ \jone\ map by Rubio et al. (1993). This map covers only a
relatively small fraction of the area mapped in \ci, but it probably
contains most of the CO emission: to obtain a \citoco\ ratio we assume
that there is no CO emission outside this region. The CO emission
associated with N27 has two components, at 114.5 and 126.9
\kmpers. For this analysis we have integrated the CO and \ci\ data
over the velocity range corresponding to the strongest component, at
$\sim114$ \kmpers.  In order to compute a ratio we have convolved the
CO \jone\ integrated intensity data to the angular resolution of the
\ci\ map. Subsequently we sampled the convolved CO map at the
positions observed in \ci, excluding those positions where \ci\
spectra were discarded.  These samples were averaged with the same
weighting scheme used for \ci.  This procedure allows us to obtain a
reliable average ratio for the N27 region, relatively insensitive to
uncertainties in the \ci\ beam size and pointing.  The resulting
intensities, integrated over the velocity interval 107---119 \kmpers,
are $\Ici\simeq0.132\pm0.027$ and $\Ico\simeq0.517\pm0.052$ \Kkmpers,
yielding a ratio $\citoco\approx0.26\pm0.06$. The corresponding
cooling ratio (the ratio of intensities in \intunits) is
$\citoco\approx19.9\pm4.5$.  The uncertainties are
$1\sigma$ and include the statistical errors as well as a 10\%
$1\sigma$ calibration uncertainty in the intensities, added in
quadrature.

Figure \ref{pdrci} compares the \citoco\ ratio in N27 to a
sample of molecular clouds associated with star-forming regions in
systems of different metallicities.  This plot suggests that there is
an increasing trend in the \citoco\ ratio with decreasing metallicity
$Z$, such that $\citoco\propto Z^{-0.5}$. The scatter around this line
is relatively large, as shown by the data for the Large
Magellanic Cloud. Theoretical expectations notwithstanding, this
scatter has made it difficult to identify a trend in past studies over
a narrower range of metallicities (e.g., Bolatto et al. 2000a).

Shown also in Fig. \ref{pdrci} is a model for carbon in unresolved PDRs
(Bolatto et al. 1999). This model consists of intensity
ratios computed over a distribution of clumps with uniform excitation,
where each spherical clump is divided in three concentric regions of
\cp, \cn, and CO. The sizes of these regions depend on the metallicity
(or the dust-to-gas ratio) of the ISM.  While necessarily crude this
model incorporates the effects of clumping, and the increase in the
physical size of the PDR for decreasing metallicity.  The growth of
the PDR is mostly caused by the diminishing attenuation of the UV
radiation, as the dust-to-gas ratio of the ISM decreases for smaller
metallicities. In this scenario, the primary reason why the \citoco\
ratio increases for decreasing metallicity is the reduction and
eventual disappearance of the CO cores in the clumps.

Are the models for the \citoco\ ratio as a function of $Z$
quantitatively consistent with the measurements?  Bolatto et al. explore two
possibilities: \cn\ produced in the growing PDR (model A), and \cn\
restricted to a region whose size does not change with metallicity
(model B).  This latter model predicts an essentially flat \citoco\
ratio ($\citoco\propto Z^{+0.1}$). A similar result (constant
$\citoco$) would be obtained if \cn\ was produced inside the CO cores
via unidentified chemical reactions.  The slope predicted by model A
in this metallicity range is slightly steeper than that suggested by
the data, $\citoco\propto Z^{-0.8}$.  Whether this difference is
actually significant is difficult to assert given the paucity of
observations and the intrinsic scatter in the data. Nevertheless,
several possibilities could explain this discrepancy: 1) an increase
in CO excitation for decreasing metallicity (i.e., hotter molecular
gas for diminishing $Z$), 2) clump-to-clump
shielding preventing the efficient photodissociation of CO, 3) a
dust-to-gas ratio not proportional to the oxygen abundance (but this
would require the dust abundance to be a slower function of $Z$,
e.g. D/G$\sim Z^{0.5}$, when observations suggest D/G$\sim Z^2$;
Lisenfeld \& Ferrara 1998), and 4) some fraction ($\sim30$\%) of the
\cn\ produced inside the CO cores instead of the PDR. We conclude
that, within the bounds imposed by the limitations of the models and
the paucity of the measurements, the observed trend for the \citoco\
ratio with metallicity shown in Fig. \ref{pdrci} agrees with
theoretical expectations for mostly photodissociation-produced \cn.

\subsection{Physical Conditions}
\label{discpcn27}

The FIR spectroscopic data can be used to determine the 
conditions prevalent in N27. Using Table \ref{tabiso} and the PDR models
by Kaufman et al. (1999) we will find the UV field, \xuv, and the gas
density, $n$. The important FIR ISM diagnostics
computed by Kaufman et al. are: 1) the \oi\ \fseno/\fsowt\ line ratio
(${\rm I_{145}/I_{63}}\sim0.051\pm0.006$, according to our
measurements), 2) the \oi\ \fsowt\ to \cii\ \fscii\ line ratio (${\rm
I_{63}/I_{158}}\sim0.42\pm0.01$), 3) the ratio of \oi\ \fsowt+\cii\
\fscii\ to the FIR continuum (using the LWS spectrum we estimate
$\Ifir\sim1.1\times10^{-2}$ \intunits, yielding ${\rm
(I_{63}+I_{158})/\Ifir}\sim0.012\pm0.002$), and 4) the \cii\ \fscii\
intensity itself. These joint four diagnostics agree very well,
placing N27 in the low density, $n\sim300$---$1000$~\percmcu, moderate
UV field, $\xuv\sim30$---$100\chi_0$, region of the $n-\xuv$ diagrams
($\chi_0$ is the interstellar UV field in the vicinity of the Sun,
$\chi_0\simeq1.6\times10^{-3}$ erg~cm$^{-2}$~s$^{-1}$; Habing 1968).
The corresponding PDR temperature in these conditions is $T_{pdr}\sim200$~K.
The FIR continuum dust temperature, derived from the LWS spectrum, is
$T_d\sim40$~K.

The density and radiation field derived using millimeter and
submillimeter lines, however, strongly disagree with the
FIR results. This lack of agreement is not surprising,
perhaps, since these lines probably originate in different regions of the
ISM. A millimeter wavelength multiline excitation analysis using several
molecular species was performed by Heikkil\"a et al. (1999), yielding
$n\sim10^{5}$~\percmcu, $T\gtrsim15$~K, and visual extinction
$\av\sim1$ for N27. The \citoco\ ratio determined in \S\ref{disccin27} 
strongly constrains the density according to the calculations
by Kaufman et al. (1999), also resulting in $n\sim10^4$---$10^5$~\percmcu\
(albeit over a much larger area).  Finally, using the \co\ \jone\
observations by Rubio et al. (1993) convolved to the FIFI beam we
compute an accurate \ciitoco\ ratio at the LWS pointing, yielding
$\ciitoco\approx7000$. This ratio alone does not constrain either
parameter very well, but implies $\xuv>100\chi_0$ and when used in
combination with the measured \citoco\ ratio results in
$\xuv\sim10^4$---$3\times10^5\chi_0$.  This is not a fair comparison,
however, since both the \ciitoco\ and \citoco\ ratios will be affected
by the low metallicity and extinction of this source. Fortunately,
Kaufman et al. (1999) provide some calculations for small $Z$ and
\av. Using their model tracks for $Z=0.1$ and $\av=1$ we find
$n\sim10^5$~\percmcu\ and $\xuv\sim30\chi_0$ for N27, based on
$\Icii/\Ifir\sim10^{-2}$ and $\Ico/\Ifir\sim10^{-6}$. These
measurements are also consistent with $n_{\rm CO}>n_{\rm C^+}$ at a
lower radiation field, if we allow the model tracks to be shifted in
the abscissa as described by the model authors.

At any rate, the diagnostics including \ci\ or molecular line
observations are inconsistent with the density $n\sim10^3$ \percmcu\
implied by the FIR transitions. Because in general the FIR lines are
generated closer to the surface of the cloud, this discrepancy may
indicate a real density gradient in the gas. Another possibility is
that the mm-submm and FIR lines originate in different regions of the
cloud, with the latter dominated by material in a more diffuse
interclump medium, or perhaps associated with the \hii\ region rather
than the molecular cloud. A third possibility is that the FIR line
ratios are significantly affected by $Z$ or \av, and using model results
computed for Galactic metallicity and large \av\ is inappropriate. If
this is the cause of the density discrepancy, however, it is
remarkable that {\em all four} FIR diagnostics agree on essentially
the same value for the density and radiation field. Independent of
which explanation is correct, we conclude that: 1) the
FIR and mm-submm measurements roughly agree on the radiation field
intensity, although it is extremely important to use models with the
appropriate $Z$ and \av, and 2) the mm-submm data systematically
indicate a density about two orders of magnitude higher than that
derived from FIR fine structure data.

\section{Summary and Conclusions}
We report new FIR and submillimeter observations of the star forming
region N27 in the Small Magellanic Cloud. Using our measured
intensities for the FIR continuum and \oi\ and \cii\ fine structure
transitions, and comparing with existing model calculations (Kaufman
et al. 1999), we derive the density and radiation field in N27 to be
$n\sim300$---1000~\percmcu\ and $\xuv\sim30$---$100\chi_0$. In 
contrast, the density derived using millimeter and submillimeter data
is two orders of magnitude higher, $n\sim10^5$~\percmcu, while the
radiation field is similar or lower, $\xuv\lesssim30\chi_0$. The cause
for this discrepancy is not clear, but we suggest that the FIR and the
mm-submm lines arise in different parcels of gas, with the former
perhaps dominated by the interclump medium or the \hii\ region.

\begin{figure}[p]
\plotone{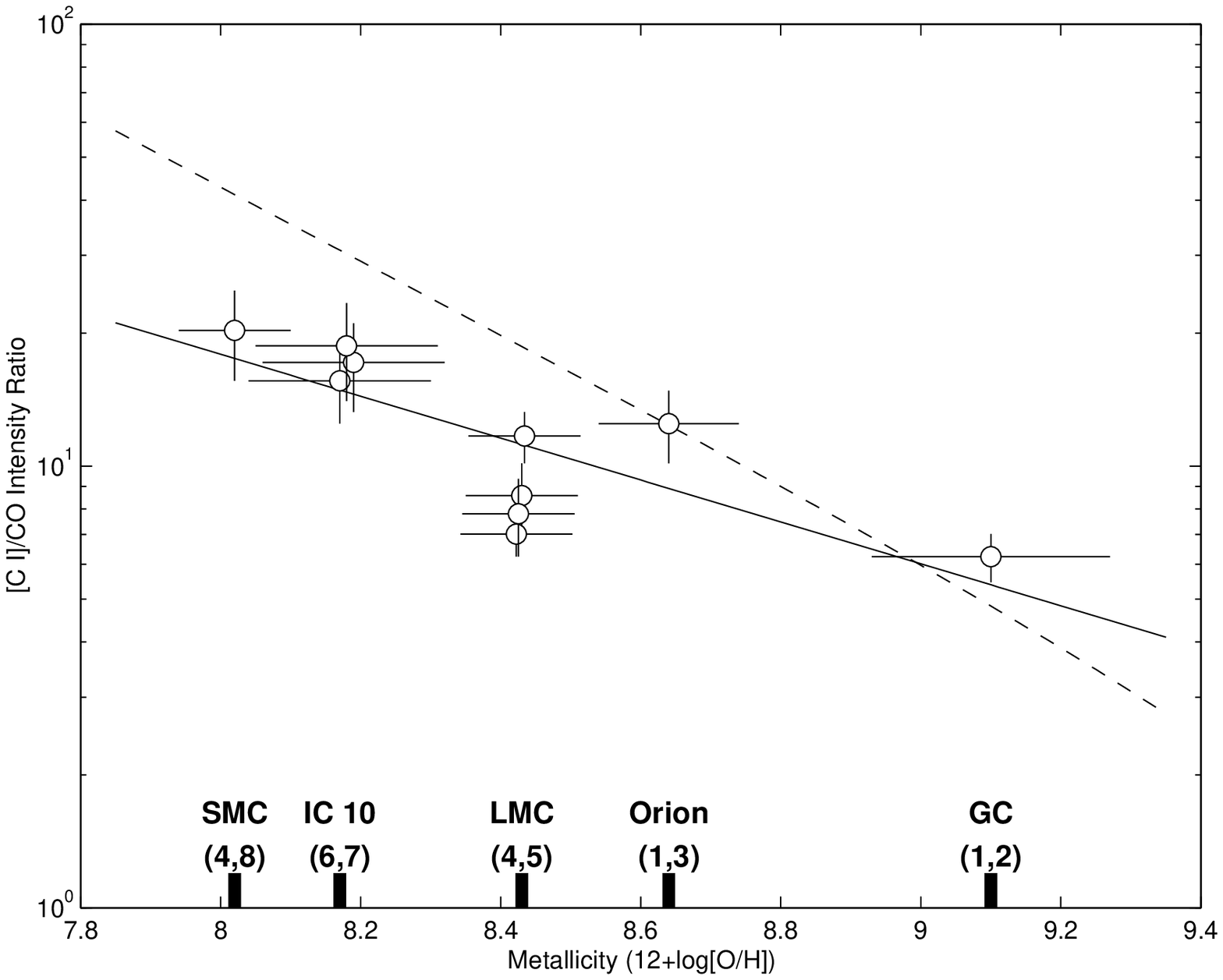} 
\figcaption[fig2.eps]{The \ci/CO line intensity ratio as a function of
metallicity, for molecular clouds in the Galactic Center (GC), the
Orion region, the LMC, IC~10, and the SMC. The continuous line is a
linear fit to the data: $\log[\citoco]=-0.47*(12+\log[{\rm
O/H}])+5.0$.  The dashed line is Bolatto et al. (1999) model A
(PDR-produced neutral carbon): $\log[\citoco]=-0.82*(12+\log[{\rm
O/H}])+8.2$.  The references for metallicity and \citoco\ ratio for
each source indicated in the figure are: (1) Afflerbach et al. 1997,
(2) Ojha et al. 2000, (3) Tauber et al. 1995, (4) Dufour 1984, (5)
Bolatto et al. 2000b, (6) Lequeux et al. 1979, (7) Bolatto et
al. 2000a, and (8) this work.
\label{pdrci}}
\end{figure}

We have detected \ci\ \fsone\ emission from the Small
Magellanic Cloud for the first time. This is the lowest metallicity
source in which submillimeter neutral carbon emission has been detected.
Low metallicity, dust-poor environments are thought to be dominated by
PDRs.  Despite the factor of $\sim5$ in metallicity spanned by N27 and
Orion, however, the \citoco\ ratio in N27 is only modestly larger than
the \citoco\ ratio in Orion. Nevertheless, when we study the ratio in
several sources spanning more than an order of magnitude in
metallicity, we find a noisy but convincing increasing trend in the
\citoco\ ratio for decreasing metallicity (Fig. \ref{pdrci}). This
trend is somewhat flatter than, but in rough agreement with, the
modelling results by Bolatto et al. (1999). The existence of an
increasing trend points to a photodissociation origin for most of neutral
carbon inside molecular clouds, in agreement with standard PDR models.

\acknowledgements This research was supported in part by grants NSF
OPP 89-20223, NSF AST 98-03065, NASA JPL 96-1513, and made use of NASA
ADS and the CDS databases. We wish to thank M\'onica Rubio for
providing the calibrated and baseline-removed CO \jone\ spectra, Frank
Israel for the FIFI \cii\ data, and Tom Bania for his useful comments.

\newpage
\begin{deluxetable}{llcrcr@{$\pm$}lcr@{$\pm$}l}
\tablewidth{0pc}
\scriptsize
\tablecaption{ISO LWS Line Intensities in N27}
\tablehead{\multicolumn{1}{c}{Specie} &
\multicolumn{1}{c}{Transition}&\multicolumn{1}{c}{Wavelength}&
\multicolumn{1}{c}{Beam Size} &\multicolumn{1}{c}{ESCF\tablenotemark{a}}&
\multicolumn{2}{c}{Line Flux} &\multicolumn{1}{c}{Continuum}&
\multicolumn{2}{c}{Line/Cont.\tablenotemark{b}}\\
&			      &\multicolumn{1}{c}{($\mu$m)}  &
\multicolumn{1}{c}{(sr)}&                                     &
\multicolumn{2}{c}{(\intunits)}&\multicolumn{1}{c}{(MJy sr$^{-1}$)}&
\multicolumn{2}{c}{}
}
\startdata
\mbox{[O~III]} & \fstwo & \phn51.815 & 
$11.8\times10^{-8}$ & 0.85 &
1.34 & $0.22\times10^{-5}$ & 191 &
0.20 & 0.03 \\
\mbox{[O~I]} & \fsowt & \phn63.184 & 
$13.0\times10^{-8}$ & 0.83 &
3.93 & $0.09\times10^{-5}$ & 237 &
0.72 & 0.02 \\
\mbox{[O~III]} & \fsone & \phn88.356 & 
$11.8\times10^{-8}$ & 0.67 &
9.64 & $0.85\times10^{-6}$ & 293 &
0.13 & 0.01 \\ 
\mbox{[O~I]} & \fseno & 145.525 & 
$6.7 \times10^{-8}$ & 0.58 &
2.01 & $0.23\times10^{-6}$ & 415 &
0.06 & 0.01 \\
\mbox{[C~II]} & \fscii & 157.741 & 
$6.7 \times10^{-8}$ & 0.58 &
9.46 & $0.08\times10^{-5}$ & 394 &
3.14 & 0.04 \\
\enddata
\tablenotetext{a}{Recommended extended source correction factor 
(ISO LWS Handbook online v1.0).}
\tablenotetext{b}{Line to continuum ratio at line center (unresolved lines, 
$\Delta\lambda=0.29$ $\mu$m for $\lambda<80$ $\mu$m. Otherwise 
$\Delta\lambda=0.60$ $\mu$m).}
\label{tabiso}
\end{deluxetable}

\newpage

\begin{references}
\reference{ACW97} Afflerbach, A., Churchwell, E., \& Werner, M. W. 1997, 
\apj, 478, 190

\reference{BJI99} Bolatto, A. D., Jackson, J. M., \& Ingalls, J. G. 1999,
\apj, 513, 275

\reference{BO00a} Bolatto, A. D., Jackson, J. M., Wilson, C. D., \& 
Moriarty-Schieven, G. 2000a, \apj, 532, 909

\reference{BO00b} Bolatto, A. D., Jackson, J. M., Israel, F. P., Zhang, X.,
\& Kim, S. 2000b, \apj, in print (astro-ph/0007149)

\reference{BO85} Bouchet, P., Lequeux, J., Maurice, E., Pr\'evot, L., \&
Pr\'evot-Burnichon, M. L. 1985, \aap, 149, 330

\reference{CL96} Clegg, P. E. et al. 1996, \aap, 315, L38

\reference{DEM76} Davies, R. D., Elliot, K. H., \& Meaburn, J. 1976, 
\memras, 81, 89

\reference{DU84} Dufour, R. J. 1984, in Structure and Evolution of the 
Magellanic Clouds, ed. S. van der Bergh \& 
K. S. de Boer (Dordrecht:Kluwer), 353

\reference{HA68} Habing, H. J. 1968, \bain, 19, 421

\reference{HE99} Heikkil\"a, A., Johansson, L. E. B., \& Olofsson, H. 1999,
\aap, 344, 817

\reference{HE56} Henize, K. G. 1956, \apjs, 2, 315

\reference{IS93} Israel, F. P. et al. 1993, \aap, 276, 25

\reference{IM93} Israel, F. P., \& Maloney, P. R. 1993, in New Aspects
of Magellanic Cloud Research, eds. B. Baschek, G. Klare and J. Lequeux
(Berlin:Springer-Verlag), 44

\reference{KA99} Kaufman, M. J., Wolfire, M. G., Hollenbach, D. J., \& Luhman,
M. L. 1999, \apj, 527, 795

\reference{LS94} Laney, C. D., \& Stobie, R. S. 1994, \mnras, 266, 441

\reference{LE79} Lequeux, J., Peimbert, M., Rayo, J. F., Serrano, A., 
\& Torres-Peimbert, S. 1979, \aap, 80, 155

\reference{LF98} Lisenfeld, U., \& Ferrara, A. 1998, \apj, 496, 145

\reference{OJ00} Ojha, R. et al. 2000, \apjlett, submitted

\reference{RU93} Rubio, M. et al. 1993, \aap, 271, 1

\reference{ST97} Stark, A. A., Chamberlin, R. A., Cheng, J., Ingalls, J. G.,
\& Wright, G. 1997, Rev. Sci. Inst., 68 (5), 2200

\reference{SD95} Sternberg, A., \& Dalgarno, A. 1995, \apjs, 99, 565 

\reference{SW96} Swinyard, B. M. et al. 1996, \aap, 315, L43

\reference{TH85a} Tielens, A. G. G. M., \& Hollenbach, D. 1985, \apj, 291, 722

\end{references}
\end{document}